\documentclass[twocolumn,showpacs]{revtex4}
\usepackage{psfig}
\begin{document}
\title{Systematic Regge theory analysis of omega photoproduction}
\author{A. Sibirtsev$^1$, K. Tsushima$^{1,2,3}$ and S. Krewald$^{1,2}$}
\affiliation{$^1$Institut f\"ur Kernphysik, Forschungszentrum J\"ulich,
D-52425 J\"ulich \\
$^{2}$Special Research Center for the Subatomic 
Structure of Matter (CSSM)  and Department of 
Physics and Mathematical Physics, 
University of Adelaide, SA 5005, Australia \\
$^3$ Department of Physics and Astronomy, University
of Georgia, Athens, GA 30602 USA \\ }

\begin{abstract}
Systematic analysis of available data for $\omega$-meson photoproduction
is given in frame of Regge theory. At photon energies above
20~GeV the $\gamma{+}p{\to}\omega{+}p$ reaction is entirely dominated 
by Pomeron exchange.  However, it was found that Pomeron exchange
model can not reproduce the  $\gamma{+}p{\to}\rho{+}p$
and $\gamma{+}p{\to}\omega{+}p$ data at high energies
simultaneously with the same set of parameters.
The comparison between $\rho$ and $\omega$ data 
indicates a large room for meson exchange contribution to $\omega$-meson 
photoproduction at low energies. It was found that at low energies the 
dominant contribution comes from $\pi$ and $f_2$-meson 
exchanges. There is smooth transition between the meson exchange model at 
low energies and Regge theory at high energies.
\end{abstract}

\pacs{13.60.Le; 13.88.+e; 14.20.Gk; 25.20.Lj} 
\maketitle
\section{Introduction}
The vector meson photoproduction at high energies is traditionally
discussed in terms of Regge theory. The most recent systematic
theoretical analysis~\cite{Donnachie1,Donnachie2,Laget1} 
confirmed that at high energies the photoproduction  
of $\rho$- and $\phi$-meson can be well described by soft 
Pomeron and meson Regge trajectories. 

ZEUS data~\cite{Breitweg} at high energies 
on $\rho$-meson photoproduction at $|t|$ above 
$\simeq$0.4~GeV$^2$ need  additional contribution from  hard Pomeron 
exchange~\cite{Donnachie3}. Furthermore, very recent CLAS data on
$\rho$-~\cite{Battaglieri} and $\phi$-meson~\cite{Anciant}
photoproduction at $E_\gamma{\simeq}3.{\div}4$~GeV can be well
explained by Regge theory at low momentum transfer 
$|t|{\le}1$~GeV$^2$. 

The Regge theory calculations~\cite{Donnachie2,Laget2}  for
$\rho$-meson photoproduction at low $|t|$ indicates that at low energies
the dominant contribution comes from  $f_2$-meson
exchange, while at high energies it is due to Pomeron exchange.
The $\phi$- and $J/\Psi$-meson photoproduction at small momentum transfers
are dominated  by Pomeron exchange
because of the $s{\bar s}$ and $c{\bar c}$ structure of 
these mesons~\cite{Sibirtsev9}, respectively.

At backward angles,  where  $|u|$ is small, the $\rho$-meson
photoproduction is dominated by  exchange of nucleon and $\Delta$ Regge 
trajectories in the $u$ channel. The backward $\phi$-meson
photoproduction is due to the $u$-channel nucleon exchange.

Moreover, the recent data~\cite{Battaglieri,Anciant} on 
$\rho$ and $\phi$-meson photoproduction at both large $|t|{>}1$~GeV$^2$
and large $|u|$ had been interpreted as due to the hard scattering 
between the photon and the quarks in the nucleon.

While $\rho$ and $\phi$-meson photoproduction were systematically studied
within the Regge theory, the $\omega$-meson photoproduction data have
been analyzed only selectively. As is shown in Refs.~\cite{Laget1,Laget2} the 
calculation with the inclusion of $\pi$, $f_2$ and Pomeron trajectory 
exchanges underestimate experimental data on total 
$\gamma{+}p{\to}\omega {+}p$ cross section at $E_\gamma{\ge}$50~GeV.
It is important that at these high energies  the photoproduction is 
dominated by Pomeron exchange and it is strongly 
believed~\cite{Donnachie1,Donnachie2} that the Regge theory is able
to describe the total cross section for the $\omega$ as
well as for the $\rho$- and $\phi$-meson photoproduction. 
 
To get a overview about the Regge theory for $\omega$-meson photoproduction
we perform a systematic analysis of available data.

\section{The Pomeron exchange}
The mechanism responsible for elastic vector meson photoproduction
at high energies was originally identified as Pomeron 
exchange~\cite{Donnachie1,Donnachie2} within the 
phenomenological Regge model~\cite{Irving}. Recently, 
there was an apparent development in the description of Pomeron 
exchange  in terms of the quark and gluon degrees of freedom of 
QCD~\cite{Pichowsky} as an underlying theory of strong interactions. 

The concept of the model is based~\cite{Donnachie5} on that 
a photon fluctuates into the vector meson, and interacts with 
the quark, among those confined in the vector meson and the nucleon 
through the Pomeron exchange. 
The  structure of the quark-nucleon Pomeron exchange
interaction and the most general expression for the  amplitude ${\cal T}_P$ 
for the vector meson production by virtual photon were given 
by Donnachie and Landshoff~\cite{Donnachie5} as
\begin{eqnarray}
{\cal T}_{P_1}{=}3 i \, F_1(t)\, \frac{8 \sqrt{6}\, m_q e_q f_V \beta_0^2}
{Q^2+4m_q^2-t}\,  (\varepsilon{\cdot}\varepsilon_V)
 \left( \frac{s}{s_0}
\right)^{\alpha_{P_1}(t)-1} \nonumber \\ 
\times \exp[\frac{-i\pi 
(\alpha_{P_1}(t){-}1)}{2}] \,\,
\frac{\mu_0^2}{2\mu_0^2+Q^2+4m_q^2-t},
\label{pom}
\end{eqnarray}
where the proton isoscalar electromagnetic (EM) form factor $F_1(t)$ 
is given by
\begin{equation}
F_1(t)=\frac{4m_p^2-2.8t}{4m_p^2-t} \, \frac{1}{(1-t/t_0)^2},
\label{form1}
\end{equation}
where $m_p$ is the proton mass and $t_0$=0.7~GeV$^2$. Furthermore,
$s$ is invariant collision energy squared, and $Q^2$ denotes the squared 
mass of the virtual photon and $t$ is the four momentum transfer squared 
between the photon and vector meson. In Eq.(\ref{pom}) $e_q$ and $m_q$ 
are the quark effective charge and mass, respectively, while
$\varepsilon$ and $\varepsilon_V$ are the polarization vectors of 
the virtual photon and vector meson, respectively. Finally $f_V$ is the
vector meson radiative decay coupling constant given by $V{\to}e^+e^-$ decay
width. 

\begin{figure}[b]
\vspace*{-5mm}
\hspace*{-3.mm}\psfig{file=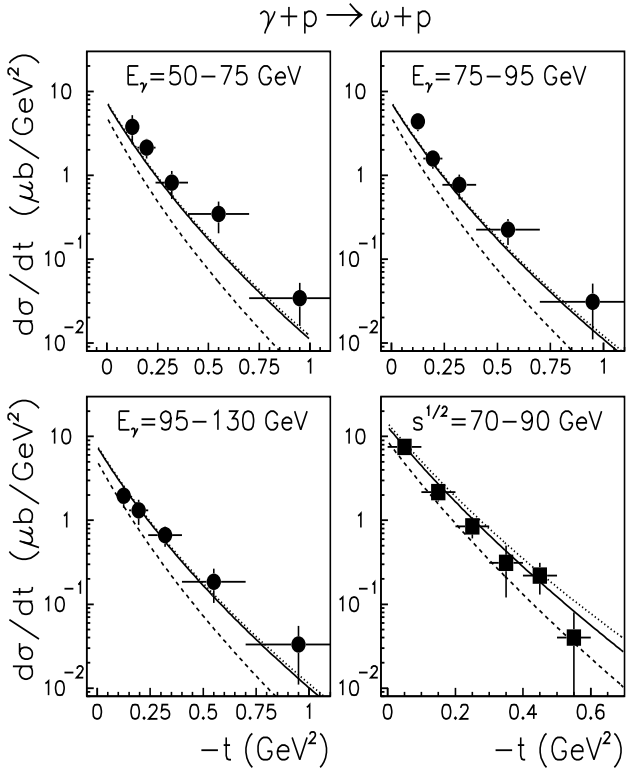,width=9.6cm,height=11.cm}
\vspace*{-9mm}
\caption[]{Differential cross section for $\gamma{+}p{\to}\omega{+}p$ as
a function of four momentum transfer squared $t$, at different photon
energies $E_\gamma$, or invariant collision energy $\sqrt{s}$.
The circles show experimental results from FNAL~\cite{Breakstone},
while the squares are the data from DESY ZEUS~\cite{Derrick}.
The lines show the Regge model calculations with Pomeron exchange 
only: the dashed- by Eq.(\ref{pom1}) with fixed 
parameter $\beta_0$=2.0~GeV$^{-1}$ \cite{Donnachie6,Donnachie7,Laget12}, 
the solid- by Eq.(\ref{phen}) with soft Pomeron exchange parameters 
adjusted by the $\gamma{+}p{\to}\omega{+}p$ data and the dotted- 
with both soft and hard Pomeron exchanges.}
\label{reg5}
\end{figure}

Free phenomenological parameters  are 
$\beta_0$ and $\mu_0$. The parameter $\beta_0$ determines 
the strength of the effective coupling of the Pomeron to quark,
while the parameter $\mu_0$ accounts that the coupling is not 
point-like and it is dressed with the form factor given by the last term
of Eq.(\ref{pom}). The experimental values of $\beta_0$ and 
$\mu_0$  were evaluated from the high energy elastic and inelastic
scattering at small $|t|$, and are given 
as~\cite{Donnachie6,Donnachie7,Laget12}
\begin{equation}
\beta_0=2.0~\mbox{GeV}^{-1}\!\!, \hspace{7mm} \mu_0=1.1~\mbox{GeV}.
\label{param}
\end{equation}

Furthermore, the Pomeron trajectory $\alpha_{P_1}(t)$ is given 
by~\cite{Donnachie4}
\begin{equation}
\alpha_{P_1}(t)=1+\epsilon+\alpha^\prime_{P_1} t,
\label{trak1}
\end{equation}
with $\epsilon$=0.08 and $\alpha^\prime_{P_1}$=0.25~GeV$^{-2}$. 
The constant $s_0$ is not well determined theoretically, however 
within the dual model prescription~\cite{Veneziano} it can be taken 
as $s_0{=}1{/}\alpha^\prime_{P_1}$.

Finally, the differential $\gamma{+}p{\to}V{+}p$ cross section 
$d\sigma{/}dt$ due to  Pomeron exchange is given for a real photon, 
\begin{eqnarray}
\frac{d\sigma}{dt}=\frac{81\, m_V^3 \, \beta_0^4 \,\, \mu_0^4 \,\,
\Gamma_{e^+e^-}}
{\pi \alpha} \, \left( \frac{s}{s_0}
\right)^{2\alpha_{P_1}(t)-2}\nonumber \\ \times \frac{F_1^2(t)}
{(m_V^2-t)^2\,\, (2\mu_0^2+m_V^2-t)^2},
\label{pom1}
\end{eqnarray}
where $m_V{=}2m_q$ is the vector meson mass, $\Gamma_{e^+e^-}$ is the
partial $V{\to}e^+e^-$ decay width and $\alpha$ is electromagnetic 
coupling constant.

The dashed lines in Fig.\ref{reg5} show the contribution from Pomeron
exchange to the differential $\gamma{+}p{\to}\omega{+}p$ cross section as
a function of four momentum transfer squared $t$. The circles in
Fig.\ref{reg5} indicate the FNAL data~\cite{Breakstone} collected
at photon energies $50{\le}E_\gamma{\le}130$~GeV, while the
squares show the DESY ZEUS data~\cite{Derrick} at invariant 
collision energy $70{\le}\sqrt{s}{\le}90$~GeV. Obviously the 
calculations by Eq.(\ref{pom1}) with $\beta_0$=2.0~GeV$^{-1}$
underestimate the data by a factor 1.9.

Most recently the parameters of the Pomeron exchange were 
readjusted~\cite{Sibirtsev0} to $\gamma{+}p{\to}\omega{+}p$ data,
which can be well reproduced with the coupling constant, 
$\beta_0{=}2.35$~GeV$^{-1}$. We also note that the same value 
of the coupling constant was deduced~\cite {Pichowsky} from an 
analysis of $\pi{+}p$ elastic scattering data at high energies.

The Pomeron parameters were also fixed~\cite{Donnachie2,Donnachie8} 
with the $\gamma{+}p{\to}\rho^0{+}p$ data.
It was  assumed that at high energies the contribution from Pomeron exchange
to the total $\rho{+}p$ cross section is the same as to    
the averaged total $\pi^+{+}p$ and $\pi^-{+}p$ cross sections.
Then the vector dominance model through the $\gamma\rho$ conversion
was applied. Finally, the phenomenological Pomeron exchange 
amplitude was given  as
\begin{equation}
{\cal T}_{P_1}{=}iA_{P_1} F_1(t)\,G(t)\left( \frac{s}{s_0}
\right)^{\alpha_{P_1}(t){-}1} \!\!\!\!\!\!\!\!\!\!\! \exp[\frac{-i\pi 
(\alpha_{P_1}(t){-}1)}{2}],
\label{phen}
\end{equation}
with the proton EM form factor of Eq.(\ref{form1}). The amplitude
is normalized such that $d\sigma{/}dt{=}|{\cal T}|^2$ in $\mu$bGeV$^{-2}$.
The Pomeron trajectory was defined by Eq.(\ref{trak1}), while the 
$\gamma\rho$ vertex function $G(t)$ was fitted~\cite{Donnachie2} 
to the $\gamma{+}p{\to}\rho^0{+}p$ data as
\begin{equation}
G(t)=\frac{1}{1-t/0.71}.
\label{gt}
\end{equation}
The normalization constant of Eq.(\ref{phen}) was fitted~\cite{Donnachie2}
to the data, and extracted value is $A_{P_1}$=0.06~fm/GeV for 
the $\rho$-meson photoproduction. 

It is clear that the $t$-dependence 
of Eq.(\ref{phen}) differs from that given by Eq.(\ref{pom}).
However, at $t{=}0$ the relation between the normalization 
constant $A_{P_1}$ and the parameters $\beta_0$ and $\mu_0$ of  
Eq.(\ref{phen}) can be written as
\begin{equation}
\beta_0^4=\frac{\pi \, \alpha \, m_V}{81 \, \mu_0^4\, \Gamma_{e^+e^-}}\,
(2\mu_0^2+m_V^2)^2 \, A_{P_1}^2.
\label{relat}
\end{equation}

\begin{figure}[t]
\vspace*{-2mm}
\hspace*{-2.mm}\psfig{file=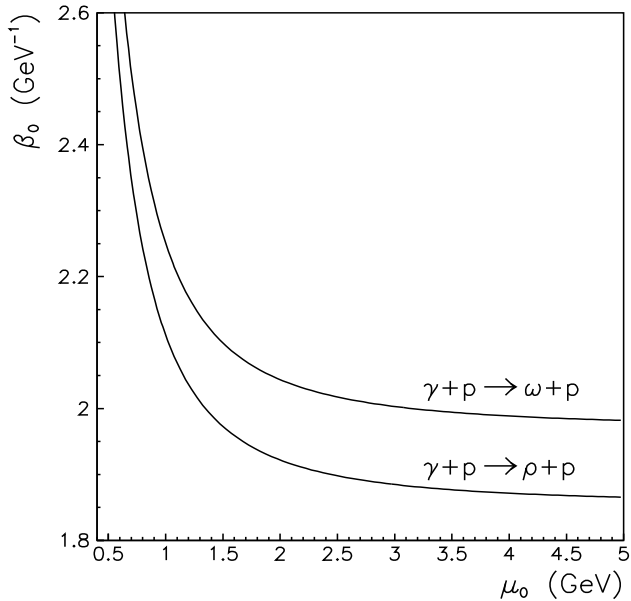,width=9.4cm,height=9.cm}
\vspace*{-9mm}
\caption[]{The solution for $\mu_0$ and $\beta_0$
parameters evaluated at $t{=}0$  by Eq.(\ref{relat}) from 
phenomenological analysis  of $\gamma{+}p{\to}\rho{+}p$
and $\gamma{+}p{\to}\omega{+}p$ data.} 
\label{reg6}
\end{figure}

Fig.~\ref{reg6} shows the solution for $\mu_0$ and $\beta_0$
parameters evaluated  by Eq.(\ref{relat}) from 
$A_{P_1}$=0.06~fm/GeV. The parameters given by 
Eq.(\ref{param}) are in reasonable agreement with phenomenological 
analysis~\cite{Donnachie2} of $\rho$-meson photoproduction.

To reproduce the $\gamma{+}p{\to}\omega{+}p$ data we adjust
the normalization constant in Eq.(\ref{phen}) as
$A_{P_1}{=}$0.02~fm/GeV.
The solid lines in Fig.~\ref{reg5} show the calculations using 
the phenomenological Pomeron  exchange amplitude of Eq.(\ref{phen}).
The calculations reproduce both the absolute value
and $t$-dependence of the differential cross section, 
$\gamma{+}p{\to}\omega{+}p$, at high photon energies.

Fig.~\ref{reg6} also shows the solution for $\mu_0$ and $\beta_0$
parameters evaluated  by Eq.(\ref{relat}) from 
$\gamma{+}p{\to}\omega{+}p$ data. It is clear that Pomeron exchange
model can not reproduce the  $\gamma{+}p{\to}\rho{+}p$
and $\gamma{+}p{\to}\omega{+}p$ data simultaneously with the same set
of parameters, $\mu_0$ and $\beta_0$. This result is in agreement with
our finding reported in Ref.~\cite{Sibirtsev0}.

\begin{figure}[b]
\vspace*{-4mm}
\hspace*{-2.mm}\psfig{file=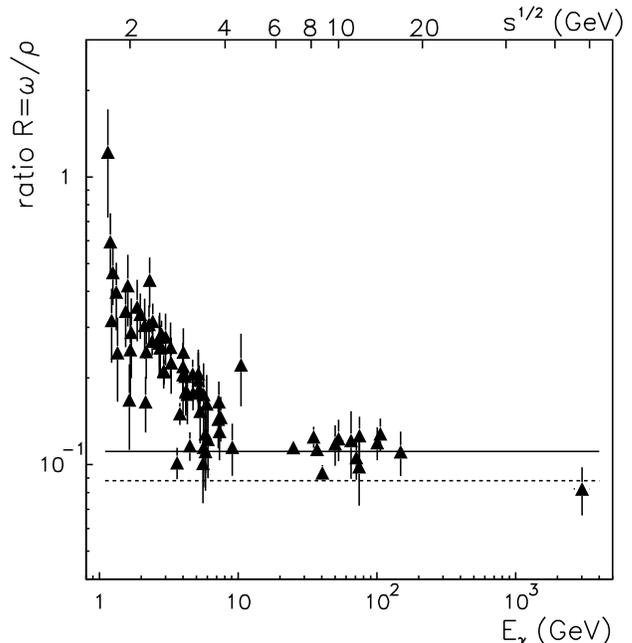,width=9.4cm,height=9.2cm}
\vspace*{-9mm}
\caption[]{The ratio of total cross sections, 
$\gamma{+}p{\to}\omega{+}p$ to 
$\gamma{+}p{\to}\rho^0{+}p$, as a function of
photon energy $E_\gamma$ (lower axis), or invariant
collision energy $\sqrt{s}$ (upper axis). The dashed line 
shows the ratio of  $\gamma\omega$ to $\gamma\rho$ coupling 
constants determined by relevant
radiative $e^+e^-$ decay widths. The solid line indicates the ratio 
expected from SU(3) symmetry.} 
\label{reg12a}
\end{figure}

Direct experimental illustration~\cite{Sibirtsev0} of this discrepancy 
is given by the ratio of the total $\gamma{+}p{\to}\omega{+}p$ to 
$\gamma{+}p{\to}\rho^0{+}p$ cross sections, which is shown
in Fig.~\ref{reg12a}. At photon energies $E_\gamma{\ge}$6~GeV it is 
consistent with SU(3) predictions and equals to 1/9. Obviously, this 
ratio is consistent with phenomenological fit given by Eq.(\ref{phen}).

The ratio of the  $\gamma{+}p{\to}\omega{+}p$ to 
$\gamma{+}p{\to}\rho^0{+}p$ cross sections from Pomeron exchange
model of Eq.(\ref{pom}) equals to the ratio  of the  $\omega{\to}e^+e^-$
and $\rho{\to}e^+e^-$ decay widths.
The experimental results on $\rho{\to}e^+e^-$
and $\omega{\to}e^+e^-$ decay widths indicate~\cite{Sibirtsev1} 
that the $\gamma\omega$ coupling is 3.4$\pm$0.2 times smaller
than that for the $\gamma\rho$. Therefore the ratio of the  
$\gamma{+}p{\to}\omega{+}p$ to $\gamma{+}p{\to}\rho^0{+}p$ 
cross sections is underestimated by $\simeq$28\% as compared with the
ratio given by the experimental $\rho{\to}e^+e^-$
and $\omega{\to}e^+e^-$ decay widths, which is shown by the
dashed line in Fig.~\ref{reg12a}.

The DESY ZEUS  data on $\rho$-meson photoproduction at 
invariant collision energy $\sqrt{s}$=71.7~GeV at 
four momentum transfer squared $0.4{\le}|t|{\le}1.6$~GeV$^2$ 
require~\cite{Donnachie2} some additional contribution, which 
may come from the hard Pomeron exchange. The hard Pomeron amplitude
is given by
\begin{equation}
{\cal T}_{P_0}{=}iA_{P_0} F_1(t)\left( \frac{s}{s_0}
\right)^{\alpha_{P_0}(t){-}1} \!\!\!\!\!\!\!\!\!\!\! \exp[\frac{-i\pi 
(\alpha_{P_0}(t){-}1)}{2}],
\label{phen1}
\end{equation}
with the proton EM form factor given by Eq.(\ref{form1}).

\begin{figure}[b]\vspace*{-5mm}
\hspace*{-4.mm}\psfig{file=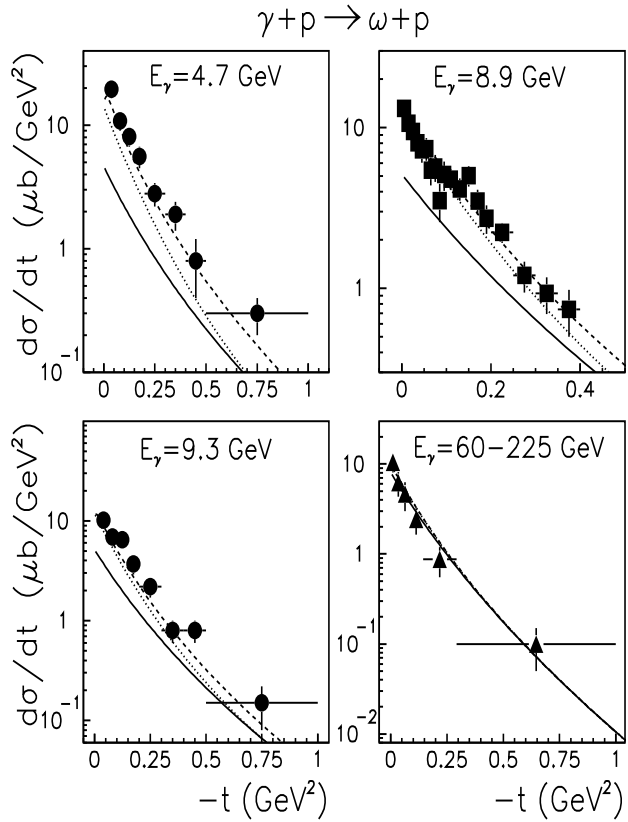,width=9.6cm,height=11.8cm}
\vspace*{-10mm}
\caption[]{Differential cross section, 
$\gamma{+}p{\to}\omega{+}p$ as
a function of four momentum transfer squared $t$ at different photon
energies $E_\gamma$.
Circles show experimental results from SLAC~\cite{Ballam},
the squares are the CORNELL data~\cite{Abramson}, while triangles
show the FNAL data~\cite{Busenitz}.
The solid lines indicate the calculations with soft and hard Pomeron 
exchanges, with parameters fitted to the 
$\omega$-meson photoproduction data. The dotted line shows the results 
with additional inclusion of $f_2$-meson exchange, while the dashed 
lines are the full model results with $\pi$, $f_2$ and Pomeron 
exchanges.}
\label{reg7}
\end{figure}

The hard Pomeron trajectory is fixed by $\rho$-meson photoproduction 
data as
\begin{equation}
\alpha_{P_0}=1.44+\alpha_{P_0}^\prime t,
\end{equation}
where $\alpha_{P_0}^\prime$=0.1~GeV$^{-2}$. Furthermore,  
$s_0{=}1{/}\alpha^\prime_{P_0}$ and $A_{P_0}{=}3.6{\cdot}10^{-4}$~fm/GeV
for $\rho$-meson photoproduction~\cite{Donnachie2}. Following the
results for soft Pomeron exchange we readjust 
$A_{P_0}{=}1.2{\cdot}10^{-4}$~fm/GeV for $\omega$-meson photoproduction.

The  $\gamma{+}p{\to}\omega{+}p$ differential cross section due to the
soft $P_1$ and hard $P_0$ Pomeron exchanges is finally given  as
\begin{equation}
\frac{d\sigma}{dt} = |{\cal T}_{P_0}+{\cal T}_{P_1}|^2,
\label{phent}
\end{equation}
and is shown in Fig.~\ref{reg5} by the dotted lines. The inclusion of 
$P_0$ exchange does not affect the calculations at low $|t|$, but
results in non negligible contribution at $70{\le}\sqrt{s}{\le}90$~GeV and
$0.4{\le}|t|$. The absence of the 
high energy $\omega$-meson photoproduction data at large $|t|$
indeed does not allow presently to clarify the role of hard 
Pomeron exchange for the $\gamma{+}p{\to}\omega{+}p$ reaction. 

\begin{figure}[t]
\hspace*{-5.mm}\psfig{file=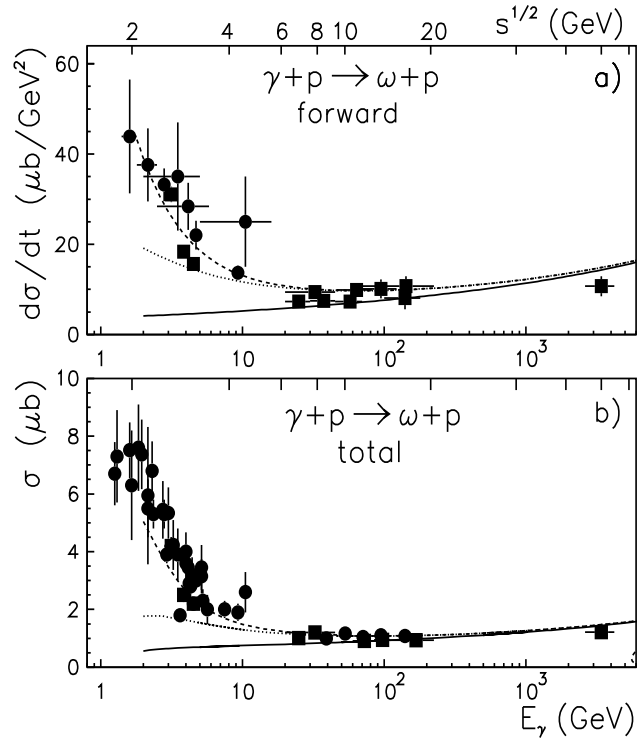,width=9.8cm,height=10.6cm}
\vspace*{-9mm}
\caption[]{Forward (extrapolated at $t{=}0$) (a) and total (b) 
$\gamma{+}p{\to}\omega{+}p$ 
cross sections as a function of photon energy $E_\gamma$ (lower axis), 
or invariant collision energy $\sqrt{s}$ (upper axis).
The solid lines indicate the calculations with soft and hard Pomeron
exchanges. The dotted lines show the results 
with additional inclusion of $f_2$-meson exchange, while the dashed 
lines are the full model results with $\pi$, $f_2$ and Pomeron 
exchanges.}
\label{reg2}
\end{figure}

With decreasing the photon energy the 
contribution from Pomeron exchange to $\omega$-meson 
photoproduction decreases. This is illustrated by Fig.~\ref{reg7}
where the experimental results~\cite{Ballam,Abramson,Busenitz}
on differential $\gamma{+}p{\to}\omega{+}p$ cross 
section are shown for the photon energies 
$E_\gamma$ from 4.7 to 225~GeV, together with soft and hard Pomeron exchange
calculations, which are indicated by the solid 
lines. 

The analysis~\cite{Donnachie2} of $\rho$-meson photoproduction
shows that at low energies significant contribution comes from
$f_2$-meson exchange. The contribution from  $f_2$-meson exchange
to the $\gamma{+}p{\to}\omega{+}p$ reaction will be analyzed in next
section. 

To summarize our analysis of Pomeron contribution
to the $\omega$-meson photoproduction is shown in Fig.~\ref{reg2}, 
for the  $\gamma{+}p{\to}\omega{+}p$ data on total 
and differential $d\sigma{/}dt$ cross section extrapolated at 
$t{=}0$, together with the Pomeron exchange calculations 
by Eq.(\ref{phent}) (the solid lines). 

The upper horizontal axis of Fig.~\ref{reg2} shows the $\gamma{+}p$ 
invariant collision energy, while the lower horizontal axis indicates the
photon energy $E_\gamma$. It is clear that the Pomeron exchange 
alone well describes the $\gamma{+}p{\to}\omega{+}p$ reaction 
at $E_\gamma$ above 20~GeV at small $|t|$ and dominates the total
$\omega$-meson photoproduction cross section. However, the Pomeron
exchange parameters for $\omega$-meson photoproduction differ from
that used~\cite{Donnachie2,Laget1} for $\gamma{+}p{\to}\rho{+}p$ reaction. 

\section{The Reggeons exchanges}
In order to reproduce the $\rho$-meson photoproduction data at 
invariant collision energies $\sqrt{s}{<}$10~GeV it is necessary 
to consider the contributions from the next to leading 
Pomeron $\rho$, $\omega$ and 
$f_2$ singularities with an intercept close to 
$\alpha(t{=}0){\simeq}0.5$. 

However the $\rho$ and $\omega$ trajectories 
can not be exchanged and it was proposed in Refs.~\cite{Donnachie2,Laget1}
that the introduction of the $f_2$ exchange may be enough to describe 
the $\gamma{+}p{\to}\omega{+}p$ data at low energies. The $f_2$ 
exchange amplitude is given by~\cite{Donnachie2,Laget1}
\begin{equation}
{\cal T}_{f}{=}iA_f\, F_1(t)\,G(t)\left( \frac{s}{s_0}
\right)^{\alpha_{f}(t){-}1} \!\!\!\!\!\!\!\!\!\!\! \exp[\frac{-i\pi 
(\alpha_{f}(t){-}1)}{2}],
\label{phen2} 
\end{equation}
with Regge trajectory
\begin{equation}
\alpha_f(t)=0.55+\alpha_f^\prime t,
\end{equation}
where $\alpha_f^\prime$=0.93~GeV$^{-2}$, $F_1(t)$ is proton 
isoscalar form factor
given by Eq.(\ref{form1}), while $G(t)$ is $\gamma\rho$ form 
factor from Eq.(\ref{gt}). Furthermore, $s_0{=}1{/}\alpha_f^\prime$ and 
parameter $A_f{=}0.159$~fm/GeV was fitted to the $\rho$-meson 
photoproduction data. It is important that the sum of the 
$f_2$ and soft and  hard Pomeron exchanges amplitude describe well the
$\rho$-meson photoproduction data on differential and total reaction
cross sections.
 
The calculations with $f_2$ and Pomeron exchanges amplitude for the 
differential $\gamma{+}p{\to}\omega{+}p$ cross section at photon energies
from 4.7 to 9.3~GeV are shown in Fig.~\ref{reg7} by the dotted lines.
We scaled the parameter $A_f$ by the factor of 3 as compared to 
the $\rho$-meson photoproduction, and performed the calculations
using the value $A_f{=}0.053$~fm/GeV for $\omega$-meson photoproduction.

The inclusion of $f_2$-meson trajectory substantially improves agreement
for the differential $\gamma{+}p{\to}\omega{+}p$ cross section at
photon energies $E_\gamma$=8.9 and 9.3~GeV. However, it is clear
that at $E_\gamma$=4.7~GeV an additional contribution is necessary
in order to reproduce the data. The dotted lines in Fig.~\ref{reg2}
show the contribution from $f_2$ and soft and hard Pomeron
exchanges to the forward and total $\omega$-photoproduction cross
sections.

\begin{figure}[t]
\hspace*{-3.mm}\psfig{file=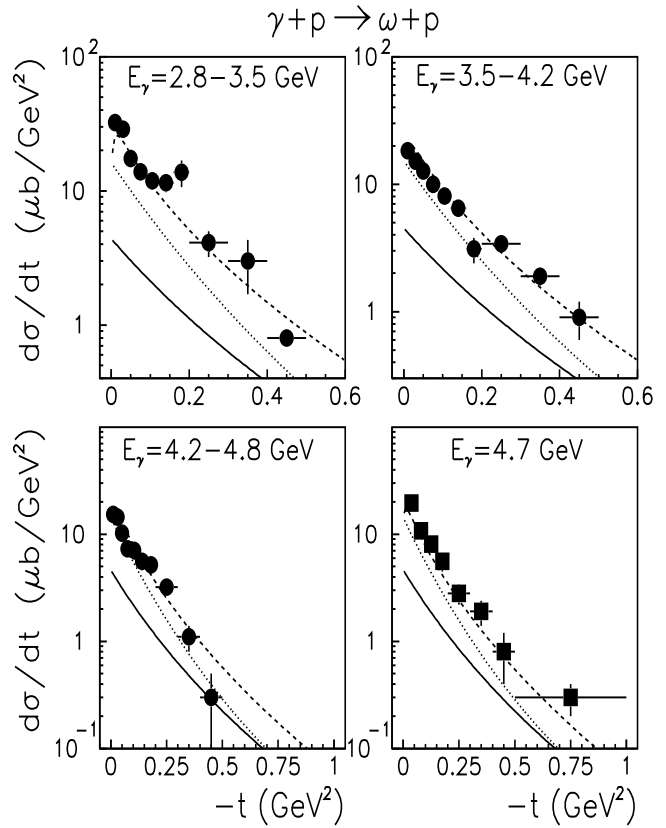,width=9.6cm,height=11.8cm}
\vspace*{-10mm}
\caption[]{Differential $\gamma{+}p{\to}\omega{+}p$ cross section as
a function of four momentum transfer squared $t$ at different photon
energies $E_\gamma$. Squares show the SLAC data~\cite{Ballam}, while
the circles are DARESBURY experimental data~\cite{Barber}.
The solid lines indicate the calculations with Pomeron exchange, 
the dotted lines show the results 
with an additional inclusion of $f_2$-meson exchange, while the dashed 
lines are the full model results with $\pi$, $f_2$ and Pomeron 
exchanges.}
\label{reg3}
\end{figure}

In case of $\omega$-meson photoproduction this contribution comes
essentially from $\pi$-meson exchange. The $\rho$-meson photoproduction 
data do not indicate much free room for the $\pi$-meson exchange since the
$\gamma\rho\pi$ coupling constant given by the $\rho{\to}\gamma\pi$
decay width $\simeq$0.1~MeV is substantially smaller
than that for the $\gamma\omega\pi$ given by $\omega{\to}\gamma\pi$
decay width of $\simeq$0.7~MeV.

This fact is as well illustrated by Fig.~\ref{reg12a}, which shows the ratio
of total $\gamma{+}p{\to}\omega{+}p$ to $\gamma{+}p{\to}\rho^0{+}p$
cross sections. Substantial enhancement of the ratio at photon energies
$E_\gamma{\le}$5~GeV might be due to $\pi$-meson exchange
contribution to $\omega$-meson photoproduction.

The $\pi$-meson exchange amplitude can be parameterized 
as~\cite{Irving1}
\begin{equation}
\!{\cal T}_\pi{=} \frac{iA_\pi m_V\sqrt{{-}t}}{t{-}m_\pi^2}\,
{\tilde G}(t)\!\left( \frac{s}{s_0}
\right)^{\alpha_\pi(t){-}1} \!\!\!\!\!\!\!\!\!\!\!\! \exp[\frac{{-}i\pi 
(\alpha_\pi(t){-}1)}{2}],\!\!
\label{phen34} 
\end{equation}
with $\pi$-meson trajectory given by
\begin{equation}
\alpha_\pi(t)=\alpha_\pi^\prime (t-m_\pi^2),
\end{equation}
where $\alpha_\pi^\prime$=0.7~GeV$^{-2}$, $m_\pi$ is $\pi$-meson mass,
$s_0{=}1{/}\alpha_\pi^\prime$. 

In the calculation, the parameter $A_\pi$=0.1~fm/GeV was 
adjusted in order to reproduce the differential 
$\gamma{+}p{\to}\omega{+}p$ cross section at photon energy 
$E_\gamma$=4.7~GeV. 

Furthermore, ${\tilde G}(t)$ denotes the 
form factor in the $\pi{NN}$ vertex fitted to the data and given by, 
\begin{equation}
{\tilde G}(t) =\frac{1}{1-t/1.3}.
\end{equation}

Now, the dashed lines in Fig.~\ref{reg7} shows the calculations with 
$\pi$, $f_2$ and soft and hard Pomeron exchanges in comparison with
experimental results on differential $\gamma{+}p{\to}\omega{+}p$ cross 
section as a function of four momentum transfer squared. The full model
results well reproduce the data. The dashed lines in Fig.~\ref{reg2}
also show the full model calculations for the forward and total
$\omega$-meson photoproduction. Agreement between the experimental 
results and the Regge theory calculation is excellent at 
$E_\gamma$ from $\simeq$5~GeV and up to DESY ZEUS energies. Moreover,
the contribution from $\pi$-meson exchange strongly decreases 
with energy.

Next we investigate down to which photon energy Regge theory can 
reproduce available data on differential cross section for 
$\omega$-meson photoproduction.

\begin{figure}[b]
\vspace*{-5mm}
\hspace*{-2.mm}\psfig{file=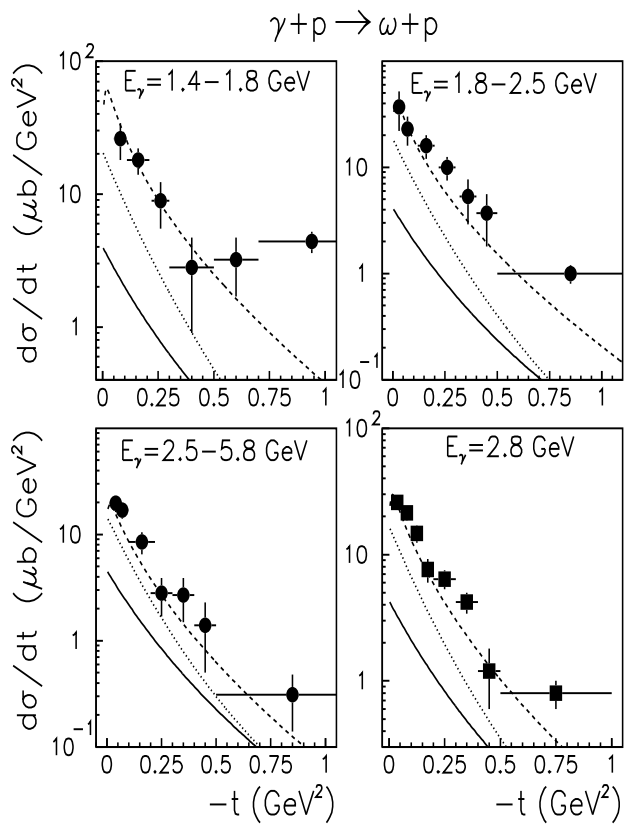,width=9.4cm,height=11.7cm}
\vspace*{-10mm}
\caption[]{Differential $\gamma{+}p{\to}\omega{+}p$ cross section as
a function of four momentum transfer squared $t$ at different photon
energies $E_\gamma$. Squares are the SLAC data~\cite{Ballam}, while
the circles show DESY experimental results~\cite{Aachen1}.
The solid lines indicate the calculations with Pomeron exchange, 
the dotted lines show the results 
with additional inclusion of $f_2$-meson exchange, while the dashed 
line are the full model results with $\pi$, $f_2$ and Pomeron 
exchanges.}
\label{reg4}
\end{figure}

Fig.~\ref{reg3} shows the
calculations together with the experimental results~\cite{Ballam,Barber}
collected at photon energies from 2.8 to 4.8~GeV. The contribution
from Pomeron exchange is indicated by the solid lines, the sum of $f_2$
and Pomeron exchanges is shown by the dotted lines. The
dashed lines in Fig.~\ref{reg3} show the full model results with
inclusion of $\pi$, $f_2´$ and Pomeron exchanges and well describe the
data even at very low photon energies. It is expected that $\pi$-meson 
exchange becomes dominant with decreasing $E_\gamma$ and at small
$|t|$.

Fig.\ref{reg4} shows data~\cite{Ballam,Aachen1} on differential 
$\gamma{+}p{\to}\omega{+}p$ cross section as a function of 
four momentum transfer squared 
collected at different photon energies within the
range $1.4{\le}E_\gamma{\le}5.8$~GeV. The contribution from Pomeron
exchange is shown by the solid lines and it is almost negligible.
The sum of $\pi$, $f_2$ and Pomeron exchanges are shown by the dashed 
lines and can describe well the data. 
Again, the $\omega$-meson photoproduction here is dominated by 
the $\pi$-meson exchange. 
 
\section{Conclusion}
Analysis of the $\omega$-meson photoproduction data and comparison
to Regge theory calculations show that at photon energies above
20~GeV the reaction is entirely dominated by Pomeron exchange.
However, we found that Pomeron exchange
model can not reproduce the  $\gamma{+}p{\to}\rho{+}p$
and $\gamma{+}p{\to}\omega{+}p$ data simultaneously with the same set
of parameters.

\begin{figure}[t]
\vspace*{3mm}
\hspace*{-4mm}\psfig{file=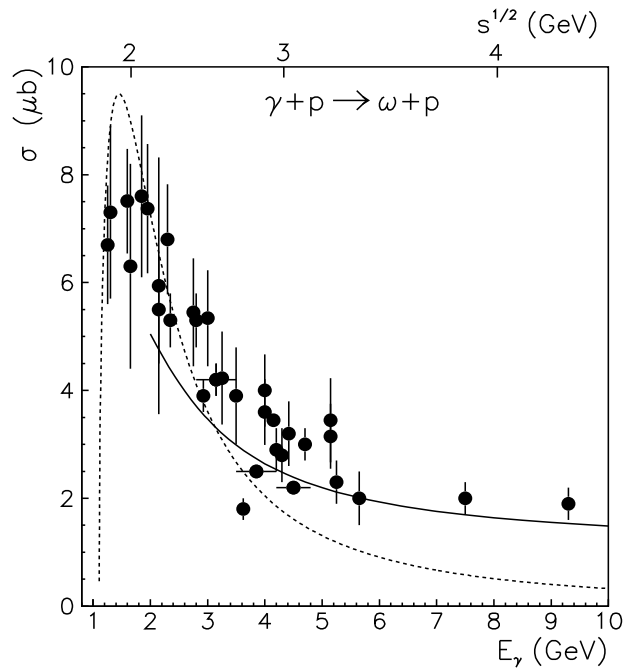,width=9.4cm,height=9.2cm}
\vspace*{-9mm}
\caption[]{Total $\gamma{+}p{\to}\omega{+}p$ cross section as
a function of  photon energy $E_\gamma$.
The solid lines indicate the calculations with Regge theory, 
while the dashed line show meson exchange model results~\cite{Sibirtsev2}.}
\label{reg9}
\end{figure}

At $E_\gamma{<}5$~GeV the dominant contribution to 
$\gamma{+}p{\to}\omega{+}p$ reaction comes from $\pi$ and 
$f_2$-meson exchanges. It is clear that appropriate parameterization 
of the form factors and coupling constants at the interaction vertices
allows us to reproduce the data even at very low energies. Instead
of parameterization of the $\pi$ and $f_2$-meson exchanges
amplitudes it is possible to describe the $\omega$-meson
photoproduction at $E_\gamma{<}5$~GeV by  meson
exchange model~\cite{Sibirtsev2}. 

Fig.~\ref{reg9} shows the data on total
$\gamma{+}p{\to}\omega{+}p$ cross section at low photon energies 
together with the Regge theory calculations and meson exchange model
results~\cite{Sibirtsev2}. Apparently there is smooth transition 
between the meson exchange model at low energies and Regge 
theory at high energies.

Finally at $E_\gamma{<}5$~GeV and for small momentum transfers the
$\omega$-meson photoproduction can be well described by both Regge
theory and by meson exchange model.


\end{document}